\documentclass[aps,preprint,prb,tighten,showpacs,graphicx,amsmath]{revtex4}

\usepackage{latexsym}
\usepackage{amssymb}

\usepackage{graphicx}
\usepackage{dcolumn}
\usepackage{bm}
\usepackage{epsfig}
\usepackage{setspace}

\begin{document}
\bibliographystyle{apsrev}

\newcommand{\Bioc}{{\it Biochemistry~}}
\newcommand{\Biom}{{\it Biomacromolecules~}}
\newcommand{\Biop}{{\it Biopolymer~}}
\newcommand{\EPJ}{{\it Eur.~Phys.~J.~}}
\newcommand{\EPL}{{\it Europhys.~Lett.~}}
\newcommand{\JCIS}{{\it J.~Coll.~Int.~Sci.~}}
\newcommand{\JPC}{{\it J.~Phys.~Chem.~}}
\newcommand{\JPCM}{{\it J.~Phys.: Condens.~Matter~}}
\newcommand{\JCP}{{\it J.~Chem.~Phys.~}}
\newcommand{\Macro}{{\it Macromol.~}}
\newcommand{\MCS}{{\it Macromol. Chem. Phys.~}}
\newcommand{\MP}{{\it Mol.~Phys.~}}
\newcommand{\PR}{{\it Phys.~Rev.~}}
\newcommand{\PRL}{{\it Phys.~Rev.~Lett.~}}

\title{Effective Electrostatic Interactions in Solutions of \\
Polyelectrolyte Stars with Rigid Rodlike Arms}
\author{Hao Wang\footnote{Present address: Dept. of Chemistry, 
Emory University, Atlanta, Georgia 30322}
and Alan R. Denton\footnote{Author to whom correspondence
should be addressed; electronic mail: {\tt alan.denton@ndsu.edu}}}
\affiliation{Department of Physics, North Dakota State University,
Fargo, North Dakota, 58105-5566}

\date{\today}

\begin{abstract}
In solutions of star-branched polyelectrolytes, electrostatic interactions
between charged arms on neighboring stars can compete with intra-star
interactions and rotational entropy to induce anisotropy in the orientational 
distribution of arms.  For model stars comprising rigid rodlike arms with 
evenly spaced charged monomers interacting via an effective screened-Coulomb 
(Yukawa) potential, we explore the influence of arm orientational anisotropy 
on effective star-star interactions.  Monte Carlo simulation and 
density-functional theory are used to compute arm orientational 
distributions and effective pair potentials between weakly charged stars.  
For comparison, a torque balance analysis is performed to obtain the 
configuration and energy of the ground state, in which the torque vanishes 
on each arm of the two-star system.  The degree of anisotropy is found to 
increase with the strength of electrostatic interactions and proximity of 
the stars.  As two stars begin to overlap, the forward arms are pushed back 
by inter-star arm-arm repulsion, but partially interdigitate due to 
rotational entropy.  At center-center separations approaching complete 
overlap, the arms relax to an isotropic distribution.  For nonoverlapping 
stars, anisotropy-induced changes in intra- and inter-star arm-arm 
interactions largely cancel and the effective pair interactions are then 
well approximated by a simple Yukawa potential, as predicted by 
linear response theory for a continuum model of isotropic stars 
[A.~R.~Denton, \PR E {\bf 67}, 11804 (2003)].  For overlapping stars, 
the effective pair interactions in the simple rigid-arm-Yukawa model 
agree closely with simulations of a molecular model that includes flexible 
arms and explicit counterions [A.~Jusufi, C.~N.~Likos, and H.~L\"owen, 
\PRL {\bf 88}, 018301 (2002); \JCP {\bf 116}, 11011 (2002)].

\end{abstract}


\maketitle

\section{Introduction}\label{Introduction}

Polyelectrolytes (PEs) are polymers that carry ionizable groups~\cite{PE}.  
In solution, small ions (counterions) dissociate, creating oppositely 
charged polyions.  Common examples of PEs are polyacrylic acid, 
sulfonated polystyrene, and biopolymers, such as proteins, DNA, and starch.  
Practical applications in the chemical, microelectronics, and pharmaceutical 
industries include water filtration membranes, direct-write 
technologies~\cite{Lewis04}, DNA-protein binding, and thin films for 
controlled drug delivery~\cite{Kakizawa02,WoodHammond05}.
When dispersed in an aqueous electrolyte, or other polar solvent, PEs 
interact electrostatically with one another and with charged surfaces.  
Bare Coulomb interactions are screened by dissociated microions 
(counterions and salt ions) distributed in and around the polyions.  
A fundamental understanding of these microscopic interactions is essential 
for predicting and controlling equilibrium and dynamical properties of 
bulk PE solutions.

Beyond linear chains, PEs can be readily synthesized with more complex 
architectures, such as stars, microgels, block-copolymer micelles, and 
dendrimers.  In this paper, we focus on star-branched PEs, comprising 
linear PE chains (arms) chemically grafted or adsorbed to a common 
microscopic core.  In sharp contrast to colloidal microspheres, 
whose hard cores are rigid and impermeable to microions, PE polyions have 
internal degrees of freedom associated with both chain conformations 
and microion distributions.  Such porous macromolecules are characterized 
by ultra-soft pair interactions~\cite{Likos01}.  The main issue addressed 
here is the extent of anisotropy in arm orientations induced by 
electrostatic interactions between neighboring stars and the implications 
for effective interactions between stars in solution.

Chain conformations in PE stars have been extensively modeled by scaling 
theory~\cite{Pincus91,Klein-Wolterink-Borisov03}, self-consistent field 
theory~\cite{Klein-Wolterink-Borisov03}, Monte Carlo 
(MC)~\cite{RogerDelsanti04} and molecular 
dynamics (MD)~\cite{Jusufi-prl02,Jusufi-jcp02} simulations.
On the experimental side, several studies have analyzed the structure of 
PE stars and spherical PE brushes in solution.
Small-angle neutron scattering~\cite{Mir95,Guenoun98,Groenewegen00-1}
and dynamic light scattering~\cite{Guo01} measurements indicate that 
the arms of highly ionized stars are almost fully stretched, a finding
supported by simulations~\cite{Jusufi-prl02,Jusufi-jcp02}.
Other neutron scattering studies, using isotopically labelled 
samples~\cite{Groenewegen00-2,Prabhu04,Prabhu05}, have directly probed 
counterion distributions, indicating a strong coupling between counterions 
and PE chains.  These studies, together with osmotic pressure 
measurements~\cite{Das02}, support conclusions from theoretical and 
simulation work~\cite{Pincus91,Jusufi-prl02,Jusufi-jcp02,Borisov91,Denton03} 
that a high fraction of counterions remain trapped inside the stars, 
screening electrostatic interactions between the arms.

Comparatively few studies have examined interactions between PE stars.
Recently, we applied response theory to a continuum 
model~\cite{Denton03,Wang04}, in which the density of charged monomers 
is approximated as an isotropic, continuously varying $1/r^2$ distribution,
where $r$ is the radial distance from the star center.  Assuming linear 
screening of bare Coulomb interactions by microions, this approach predicts 
screened-Coulomb (Yukawa) effective interactions between pairs of 
nonoverlapping stars.  In complementary studies, Jusufi, Likos, and 
L\"owen~\cite{Jusufi-prl02,Jusufi-jcp02} used simulation and variational 
theory to calculate effective pair interactions between overlapping stars 
in a molecular model that includes flexible PE chains and explicit counterions. 

Motivated by recent simulations~\cite{Jusufi-prl02,Jusufi-jcp02} and
experiments~\cite{Mir95,Guenoun98,Groenewegen00-1,Guo01}, we explore here 
a relatively simple model of PE stars consisting of rigid rodlike arms, 
lacking bend and twist flexibility, and carrying charged beads that 
interact via screened-Coulomb pair potentials.
The intrinsic rigidity of PE chains is enhanced in star architecture by 
electrostatic and solvent-mediated excluded-volume repulsions between arms.  
For example, DNA and some synthetic stiff PE chains can be assembled into 
stars with relatively inflexible arms~\cite{WangSeeman91,WangRusso01,
HeuerArcher03,Brodowski96}.  The rodlike model, while directly relevant to 
such systems, also represents a reference model for stars with 
semiflexible arms.

The outline of the remainder of the paper is as follows.  
Section~\ref{Model} first explicitly defines the rigid-arm model of PE stars.  
Section~\ref{Methods} then describes three complementary methods for 
analyzing arm orientational anisotropy and effective pair potentials 
between stars: Monte Carlo simulation, torque balance analysis, and 
density-functional theory.  Section~\ref{Results} next presents and 
discusses numerical results for arm orientational distributions and 
effective pair potentials.  Most significantly, the three independent
methods yield consistent results for star-star interactions, which 
closely agree with molecular dynamics simulations of the molecular model.  
Finally, Sec.~\ref{Conclusions} summarizes and concludes.

\section{Model}\label{Model}
\begin{figure}
\centering\includegraphics[width=0.8\textwidth]{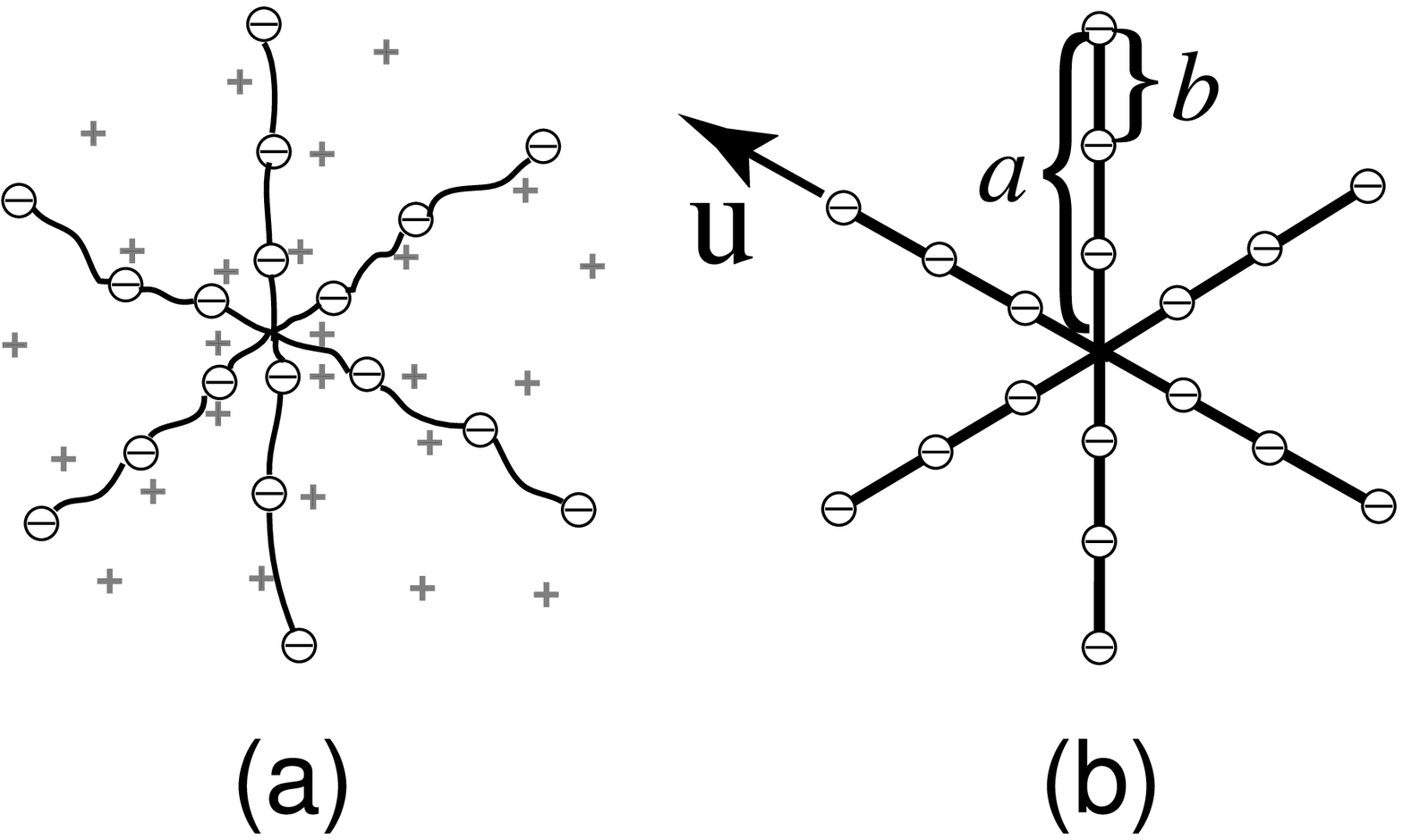}
\caption{\label{model1}{(a) Schematic drawing of PE star
and counterions. (b) Model of PE star with rigid, rodlike arms.
Integrating out degrees of freedom of the microions leads to an
effective Yukawa pair potential between pairs of charged beads
along the arms.}}
\end{figure}

We consider star-branched polyelectrolytes, each star comprising $f$ chains 
(arms) attached to a central core and freely rotating about the core.  
As depicted in Fig.~\ref{model1}, the arms are modeled as rigid rods of equal 
length $a$, each arm carrying $N_b$ charged monomers (beads) evenly spaced 
at a neighbor separation $b$ along the arm.  The microions in our model 
play two essential roles: charge renormalization and electrostatic screening.
Most of the counterions condense onto or strongly associate with the arms.  
These bound counterions reduce the bare charge on the arms, significantly 
lowering the effective charge.  The remaining mobile microions, inside and 
outside of the stars, screen the bare Coulomb electrostatic interactions.
Assuming an average effective valence $z$ per bead, the effective valence 
per star is $Z=fN_bz$.  The stars are dispersed in an electrolyte solvent, 
approximated as a dielectric continuum of spatially uniform dielectric 
constant $\epsilon$ (primitive model of electrolytes). 

In this paper, we restrict attention to ``weakly" charged stars, defined 
as stars whose arms can freely rotate relative to one another and are not 
orientationally localized.  As a simple physical criterion, a star is weakly 
charged if the increase in electrostatic energy upon rotating an arm 
half-way towards any nearest neighbor does not significantly exceed the 
typical thermal energy $k_{\rm B}T$ at temperature $T$.  
In stars not satisfying this criterion, electrostatic interactions
overwhelm Brownian motion and the arms are orientationally frozen.

To model pair interactions, we focus on two isolated stars whose centers are 
separated by a fixed displacement ${\bf R}$.  Integrating out the microion 
degrees of freedom~\cite{Denton03,Wang04} and assuming linear microion 
screening of bare Coulomb interactions, pairs of charged beads interact 
via an effective screened-Coulomb (Yukawa) potential~\cite{WangDenton}.  
The electrostatic interaction between two arms, one oriented along a unit 
vector ${\bf u}$ in a star centered at the origin, the other oriented along 
${\bf u}'$ in a star centered at ${\bf R}$, is then approximated by an 
effective pair potential
\begin{eqnarray}
\label{Varm} v_{\rm arm}({\bf u},{\bf u}';R)&=& \frac{z^2 e^2}
{\epsilon} \sum_{i=1}^{N_b} \sum_{j=1}^{N_b}\frac{\exp (-\kappa
|\mathbf{R}+ib{\bf u} -jb{\bf u}'| )}{|\mathbf{R}+ib{\bf u}-jb{\bf u}'|},
\end{eqnarray}
where $e$ is the proton charge, the summation goes over all the charged beads 
of the two arms, and $\kappa$ is the Debye screening constant (inverse Debye 
screening length).  For a solution with star number density $\rho$, 
salt ion pair number density $\rho_s$, and salt valence $z_s$, 
the screening constant is given by 
$\kappa=\sqrt{4\pi e^2(z^2 fN_b\rho+2z_s^2\rho_s)/(\epsilon k_{\rm B}T)}$.

The orientational distribution of the arms in two interacting stars is 
determined by a balance of three competing factors.  Repulsive forces
between arms on different stars exert torques that rotate the arms back, 
creating anisotropy in the arm distribution.  Opposing this backward
rotation, repulsions between arms on the same star exert torques that 
spread out the arms, favoring isotropy.  Likewise, random thermal 
(Brownian) rotational motion acts to even out the arm distribution.  
To calculate the equilibrium arm distribution, and the corresponding 
star-star interaction energy, we use three independent methods, 
described in the next section.

\section{Methods}\label{Methods}

\subsection{Monte Carlo Simulation}\label{Simulation}

\begin{figure}
\centering\includegraphics[width=0.5\textwidth]{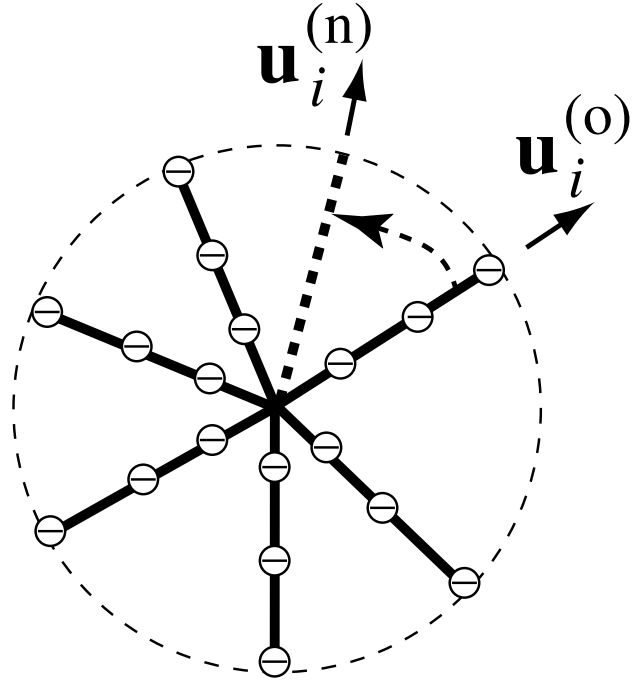}
\caption{\label{model2}{Trial rotation of the
$i^{\rm th}$ arm from its old orientation ${\bf u}^{(o)}_{i}$ to its
new orientation ${\bf u}^{(n)}_{i}$.}}
\end{figure}

For our simple rigid-arm model, Monte Carlo simulation proves to be an 
efficient method for determining the equilibrium distribution of arm 
orientations at constant temperature.  
The orientation of an arm, denoted by unit vector ${\bf u}$, is specified 
in spherical coordinates by polar and azimuthal angles ($\theta$, $\phi$).
As illustrated in Fig.~\ref{model2}, for a randomly chosen arm $i$,
a trial rotation is performed from the old orientation ${\bf u}^{(o)}_i$ to 
a new orientation ${\bf u}^{(n)}_i$.  Adopting an efficient and reliable
algorithm~\cite{FrenkelSmit}, we generate a new (trial) orientation 
according to
\begin{equation}
\label{trial} {\bf u}^{(n)}_i=\frac{{\bf u}^{(o)}_i+\gamma {\bf v}}
{|{\bf u}^{(o)}_i+\gamma {\bf v}|},
\end{equation}
where ${\bf v}$ is a unit vector with random orientation and $\gamma$ 
is a tolerance factor that determines the magnitude of the trial rotation. 
The standard Metropolis scheme gives the probability of accepting a rotation:
\begin{equation}
\label{acceptance} {\rm acc}(o \rightarrow n)=\min\left\{1,
\exp\left(-\beta[\Phi_{\rm arm}({\bf u}^{(n)}_i;R)-
\Phi_{\rm arm}({\bf u}^{(o)}_i;R)]\right)\right\},
\end{equation}
where $\beta=1/k_{\rm B}T$ and $\Phi_{\rm arm}({\bf u}_i;R)$ represents
the electrostatic energy of the $i^{\rm th}$ arm with orientation ${\bf u}_i$
in a star at center-center separation $R$ from a second star.
The latter energy can be expressed as a sum over arm-arm interactions:
\begin{equation}
\Phi_{\rm arm}({\bf u}_i;R)=\sum_{j=1\neq i}^f v_{\rm arm}
({\bf u}_i,{\bf u}_j;0)+\sum_{j=1}^{f}v_{\rm arm}({\bf u}_i,{\bf u}_j;R),
\label{armenergy} 
\end{equation}
where the first summation includes arms of the same star, excluding 
self-interaction, and the second summation includes arms of the second star.

Our simulations are initialized by centering one star at the origin, the 
other at a distance $R$ away along the $z$-axis, and assigning random 
orientations to the arms.  Next, trial rotations of the arms are performed 
in cycles, a cycle consisting of one attempted rotation of each arm, selected 
sequentially.  Following an initial equilibration stage of typically $10^3$ 
cycles, during which the energy is monitored until it reaches a plateau, 
arm orientation statistics are accumulated in the following $2\times 10^5$ 
cycles.  The average energy of the two-star system is calculated as
\begin{equation}
\label{energyMC} \Phi_{\rm eff}(R)=\frac{1}{2}\sum_{i=1}^{2f}\langle
\Phi_{\rm arm}({\bf u}_i;R)\rangle,
\end{equation}
where the summation over all arms includes intra-star and inter-star 
interaction energies and $\langle~\rangle$ denotes an ensemble average
over configurations.  Finally, the effective pair potential $v_{\rm eff}(R)$ 
between the two stars is obtained as the change in the total energy of 
the system when the two stars are brought from infinite separation 
to separation $R$:
\begin{equation}
\label{veff} v_{\rm eff}(R)=\Phi_{\rm eff}(R) - \Phi_{\rm eff}(\infty).
\end{equation}

\subsection{Torque Balance Analysis}\label{TBA}

As a check on our simulations, we also calculate the ground-state ($T=0$)
orientational distribution of arms, and the corresponding effective pair 
potential.  At zero temperature, where Brownian motion is absent, 
mechanical equilibrium is reached when the net torque on each arm 
is zero.  Electrostatic forces between arms generate torques that
drive the arms to rotate against the friction of the solvent.  Since the 
terminal angular velocity of an arm is proportional to the torque 
exerted on that arm, the equation of motion is~\cite{DoiEdwards}
\begin{equation}
\label{eqnewtonian} I_{\rm arm}\frac{\partial {\bf u}_i}{\partial t}=
-\gamma_{\rm arm}{\bf u}_i\times\nabla_{{\bf u}_i}
\Phi_{\rm arm}({\bf u}_i),
\end{equation}
where $I_{\rm arm}$ and $\gamma_{\rm arm}$ are the arm's moment of inertia 
and rotational friction coefficient, respectively.  The torque balance 
analysis amounts to finding the equilibrium configuration of the arms 
such that the right side of Eq.~(\ref{eqnewtonian}) vanishes. 

In practice, we initialize the calculation by assigning random orientations 
to the arms and then use a simple forward-difference scheme to evolve the 
system until all of the arms stop rotating.  This procedure efficiently 
generates the same final distribution as should be reached in a Monte Carlo 
simulation in which the only trial moves accepted are those that lower 
the energy.  From the final arm distribution, we compute the ground-state 
effective pair potential [from Eq.~(\ref{veff})], which represents 
a lower bound for the pair potential at nonzero temperature.

\subsection{Density-Functional Theory}\label{DFT}

An alternative to modeling the arms explicitly is to consider an 
orientational distribution function (ODF) that describes the average 
configuration of the arms.  We define the ODF $P({\bf u}; {\bf r})$ 
such that $P({\bf u}; {\bf r}){\rm d}{\bf u}$ represents the average
number of arms with orientation ${\bf u}$ in a solid angle element 
${\rm d}{\bf u}$ subtended at the center of a star at position ${\bf r}$.  
The ODF is normalized to the number of arms: 
$\int\,{\rm d}{\bf u} P({\bf u};{\bf r})=f$.  
To approximate the ODF, we propose here a 
simple density-functional theory~\cite{Oxtoby90} based on a 
mean-field approximation.

In the case that translational and rotational time scales are sufficiently 
separated that center-of-mass and orientational coordinates decouple, 
the grand potential $\Omega[\rho({\bf r}), P({\bf u}; {\bf r})]$ is a 
functional of the single-star center-of-mass number density $\rho({\bf r})$ 
and the arm ODF individually.
An arbitrary external potential $\phi_{\rm ext}({\bf r})$ uniquely 
determines $\rho({\bf r})$ and $P({\bf u}; {\bf r})$, and thus the 
grand potential~\cite{Evans79}, which can be expressed as
\begin{equation}
\Omega[\rho({\bf r}), P({\bf u}; {\bf r})]= F_{\rm id}[\rho({\bf r}), 
P({\bf u}; {\bf r})]+ F_{\rm ex}[\rho({\bf r}), P({\bf u};
{\bf r})] +\int{\rm d}{\bf r}\int{\rm d}{\bf u}\,\rho({\bf r})
P({\bf u}; {\bf r}) [\phi_{\rm ext}({\bf r})-\mu], 
\label{Omega1} 
\end{equation}
where $\mu$ is the chemical potential of the arms and $F_{\rm id}$
and $F_{\rm ex}$ are the ideal-gas and excess Helmholtz free energy 
functionals, respectively.  The ideal-gas free energy, associated with 
rotational entropy of the arms, is given exactly by
\begin{equation}
F_{\rm id}[\rho({\bf r}), P({\bf u}; {\bf r})]= k_{\rm
B}T\int{\rm d}{\bf r}\int{\rm d}{\bf u}\,\rho({\bf r}) P({\bf u};
{\bf r})[\ln(4\pi P({\bf u}; {\bf r}))-1],
\label{Fid} 
\end{equation}
translational entropy being ignored, assuming the star centers are fixed.
The excess free energy, associated with interactions between arms, 
can be formally expressed as~\cite{Evans79}
\begin{eqnarray}
F_{\rm ex}[\rho({\bf r}), P({\bf u}; {\bf r})]&=&
\frac{1}{2}\int{\rm d}{\bf r}\int{\rm d}{\bf r}'\int{\rm d}{\bf
u} \int{\rm d}{\bf u}'\int_0^1{\rm d}\lambda\,
\rho^{(2)}({\bf r},{\bf r}'; [\lambda v_{\rm arm}]) \nonumber \\
&\times&~P^{(2)}({\bf u},{\bf u}'; {\bf r},{\bf r}'; [\lambda
v_{\rm arm}]) v_{\rm arm}({\bf u},{\bf u}'; {\bf r}-{\bf r}'),
\label{Fex}
\end{eqnarray}
where $\rho^{(2)}({\bf r},{\bf r}';[v_{\rm arm}])$ is the two-star
density, proportional to the probability of finding at positions
${\bf r}$ and ${\bf r}'$ two stars whose arms interact via the 
pair potential $v_{\rm arm}({\bf u},{\bf u}'; {\bf r}-{\bf r}')$
[Eq.~(\ref{Varm})];
$P^{(2)}({\bf u},{\bf u}'; {\bf r},{\bf r}'; [v_{\rm arm}])$ is
the two-arm orientational density, proportional to the probability
of finding two arms at orientations ${\bf u}$ and ${\bf u}'$ on
stars centered at ${\bf r}$ and ${\bf r}'$; and $\lambda$ is a
coupling constant that ``turns on" the charge.

For two stars whose centers are fixed and separated by a displacement 
${\bf R}$, the single-star density can be written as
\begin{equation}
\rho({\bf r})=\delta({\bf r})+\delta({\bf r}-{\bf R}) \label{rho}
\end{equation}
and the two-star density as
\begin{equation}
\rho^{(2)}({\bf r},{\bf r}';[v_{\rm arm}])= \rho({\bf r})\rho({\bf r}').
\label{rho2}
\end{equation}
Combining Eqs.~(\ref{Omega1})-(\ref{rho2}), and setting the external potential
to zero, the grand potential simplifies to
\begin{eqnarray}
\Omega[P({\bf u})]&=&2\int P({\bf u})\left\{k_{\rm B}T[\ln(4\pi P({\bf u}))-1]
-\mu\right\} \nonumber \\
&+&\int{\rm d}{\bf u}\int{\rm d}{\bf u}'\int_0^1{\rm d}\lambda\,
\left\{P^{(2)}({\bf u},{\bf u}';0; [\lambda v_{\rm arm}]) v_{\rm
arm}({\bf u},{\bf u}';0)\right. \nonumber \\
&+&\left.P^{(2)}({\bf u},{\bf u}';R;
[\lambda v_{\rm arm}])v_{\rm arm}({\bf u},{\bf u}';R)\right\},
\label{Omega2}
\end{eqnarray}
where the factor of $2$ in the first term accounts for the two stars and
the dependence of $P({\bf u})$ on $R$ is now implicit.
In general, the two-arm density can be written as
\begin{equation}
P^{(2)}({\bf u},{\bf u}';R)= P({\bf u})P({\bf u}')
g_{\rm arm}({\bf u},{\bf u}';R), 
\label{P2}
\end{equation}
which simply defines the arm-arm pair distribution function 
$g_{\rm arm}({\bf u},{\bf u}';R)$, in such a way that,
given an arm with orientation ${\bf u}$ on a star centered at the origin,
$P({\bf u}')g_{\rm arm}({\bf u},{\bf u}';R){\rm d}{\bf u}'$ is the
average number of arms with orientation ${\bf u}'$ in solid angle 
${\rm d}{\bf u}'$ on a star centered at distance $R$.

\begin{figure}
\centering\includegraphics[width=0.8\textwidth]{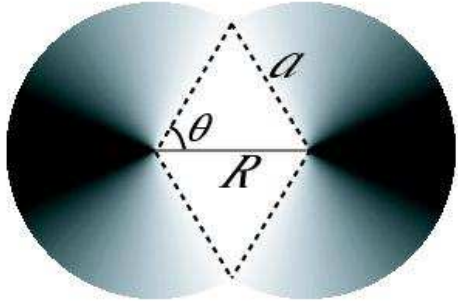}
\caption{\label{overlap}{Geometry assumed in density-functional theory 
of overlapping stars.  Arms are completely excluded from cone-shaped 
volumes (white region).  Darker shading indicates higher arm density.}}
\end{figure}

To facilitate calculations, we make two approximations.  
First, we neglect arm-arm pair correlations and take
$g_{\rm arm}({\bf u},{\bf u}';R)=1$
for both inter- and intra-star arm-arm pair distribution functions.
This mean-field approximation is based on the assumption that arm-arm 
interactions localize arms to the extent that correlations are mostly 
accounted for by the anisotropic single-arm distribution function.  
Because this approximation neglects the self-correlation hole at 
${\bf u}\simeq{\bf u}'$, it does give the wrong normalization
for intra-star correlations, $\int{\rm d}{\bf u}\int{\rm d}{\bf u}'\,
P^{(2)}({\bf u},{\bf u}';0)=f^2$,
compared with the correct result $f(f-1)$.  However, 
for multi-arm stars this is only a minor inconsistency.  In passing, we note 
that our approximation for $g_{\rm arm}({\bf u},{\bf u}';R)$ 
is analogous to that commonly applied to the pair distribution function 
$g({\bf r},{\bf r}')$ of crystals, whose particles are localized to the 
extent that the structure of the two-particle density, 
$\rho^{(2)}({\bf r},{\bf r}')=\rho({\bf r})\rho({\bf r}')g({\bf r},{\bf r}')$, 
is dominated by that of the inhomogeneous one-particle density.
As a second approximation, we assume that the arms of overlapping stars 
($R<2a$) do not interdigitate, but instead rotate backwards, being 
strictly excluded from a forward cone (Fig.~\ref{overlap}), within 
which $P({\bf u})=0$.  The validity of the no-interdigitation approximation 
is examined below in Sec.~\ref{Results}.

With the above two approximations, Eq.~(\ref{Omega2}) simplifies to
\begin{eqnarray}
\Omega[P(\mathbf{u})]&=& 2\int{\rm d}{\bf u}\,P({\bf u})\{k_{\rm B}T
[\ln(4\pi P({\bf u}))-1]-\mu\} \nonumber \\
&+&\int{\rm d}{\bf u}\int{\rm d}{\bf u}'\, P({\bf u})P({\bf u}')
[v_{\rm arm}({\bf u},{\bf u}';0)+v_{\rm arm}({\bf u},{\bf u}';R)]. 
\label{Omega3}
\end{eqnarray}
At equilibrium, $\Omega[P({\bf u})]$ is a minimum with respect to $P({\bf u})$.
Taking a functional derivative with respect to $P({\bf u})$, we obtain
\begin{equation}
\frac{1}{2}\frac{\delta\Omega[P({\bf u})]}{\delta P({\bf u})}=k_{\rm B}T
\ln(4\pi P({\bf u}))-\mu+\int{\rm d}{\bf u}'\, P({\bf u}')
[v_{\rm arm}({\bf u},{\bf u}';0)+v_{\rm arm}({\bf u},{\bf u}';R)]. 
\label{dOmega2}
\end{equation}
Setting the right side of Eq.~(\ref{dOmega2}) to zero yields an implicit
equation for $P({\bf u})$:
\begin{equation}
P({\bf u})=\frac{e^{\beta\mu}}{4\pi}\exp\left(-\beta\int{\rm d}
{\bf u}'\, P({\bf u}')[v_{\rm arm}({\bf u},{\bf u}';0)+
v_{\rm arm}({\bf u},{\bf u}';R)]\right). 
\label{P}
\end{equation}
On the right side of Eq.~(\ref{P}), the first term of the
integrand corresponds to interactions between arms in the same
star (intra-star interactions) and the second term to interactions
between arms in different stars (inter-star interactions).  Note that, 
by symmetry, the two stars are mirror images of each other reflected 
in the perpendicular plane bisecting the line joining the star centers.

Once the equilibrium orientational distribution of arms is determined, 
the energy of the two-star system -- a sum of the intra-star and inter-star 
arm interaction energies -- can be obtained from
\begin{equation}
\label{EnergyDFT} \Phi_{\rm eff}(R)=\int{\rm d}{\bf u}\int{\rm
d}{\bf u}'\,P({\bf u})P({\bf u}') [v_{\rm arm}({\bf u},{\bf
u}';0)+v_{\rm arm}({\bf u},{\bf u}';R)],
\end{equation}
which is the continuum analog of Eqs.~(\ref{armenergy}) and
(\ref{energyMC}) for the discrete model.  The effective star-star
pair interaction is finally calculated via Eq.~(\ref{veff}), taking
$P \to f/4\pi$ as $R \to \infty$.

\section{Results and Discussion}\label{Results}

Having described three methods for calculating the arm orientational 
distribution and the effective star-star pair potential, we now present
and discuss numerical results.  Our choice of parameters is restricted 
here by the defining criterion for weakly charged stars (Sec.~\ref{Model}).
In this parameter regime, where azimuthal arm-arm correlations are weak, 
we can exploit the axial symmetry of the ODF with respect to the line 
joining the centers of the two stars.  Thus, the ODF is treated as 
a function of only the polar angle: $P({\bf u})=P(\theta)$.
We have checked and confirmed this assumed symmetry in our MC simulations.  

\begin{figure}
\centering\includegraphics[width=0.8\textwidth]{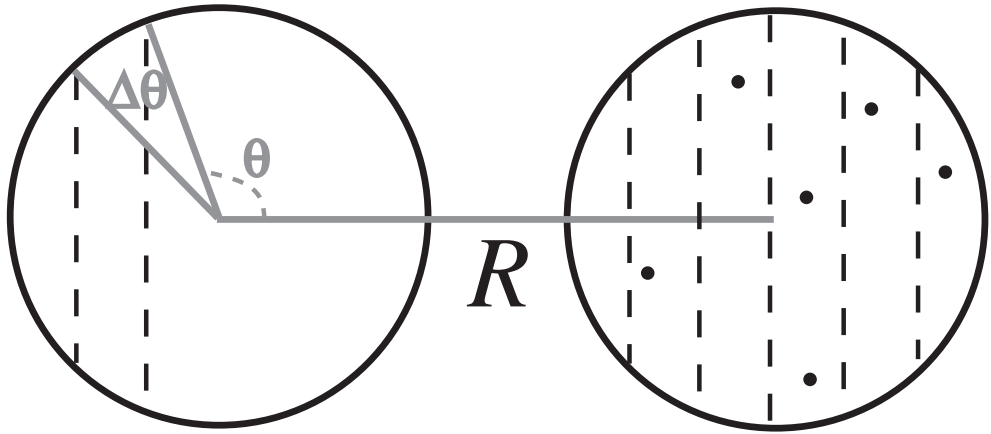}
\caption{\label{model3}{Illustration of Monte Carlo data analysis.
Dashed lines represent slices of spheres perpendicular to center-center 
line between stars.  Points indicate positions of arm tips at various 
polar angles $\theta$, each slice subtending an angle $\Delta\theta$.  
Statistical analysis of configurations yields the arm orientational 
distribution function.}}
\end{figure}

To extract $P(\theta)$ from the simulations, we bin the arm orientations 
in a manner illustrated in Fig.~\ref{model3}.  Each sphere is partitioned 
into slices perpendicular to the line joining the star centers.  
Allowing each slice to subtend the same angle $\Delta\theta$, the average
number of arms in the $k^{\rm th}$ slice centered at angle $\theta_{k}$ is 
$2\pi P(\theta_{k})\sin\theta_{k}\Delta\theta$.  The same number can also 
be expressed as $fN_{k}/\sum_{k=1}^{M}N_{k}$, where $N_{k}$ is the number 
of occurrences (accepted trial moves) of an arm in the $k^{\rm th}$ slice 
and $M$ is the total number of slices.  Equating the two expressions, 
\begin{equation}
P(\theta_k)=\frac{fN_k}{2\pi\sin\theta_k\Delta\theta\sum_{k=1}^M N_k},
\label{DFTeqMC} 
\end{equation}
which is correctly normalized to the number of arms $f$.

To compute the corresponding DFT approximation for $P(\theta)$, we solve 
Eq.~(\ref{P}) iteratively, computing the arm-arm pair potentials 
$v_{\rm arm}({\bf u},{\bf u}';R)$ from Eq.~(\ref{Varm}).  In practice, 
we construct a $60 \times 60$ grid on a sphere, with 60 equi-spaced 
longitudinal lines in the azimuthal coordinate ($0<\phi<2\pi$) and 
60 equi-spaced latitudinal lines in the polar coordinate ($0<\theta<\pi$).  
Starting from an initial normalized isotropic distribution, $P=f/4\pi$, 
we solve Eq.~(\ref{P}) self-consistently by numerical iteration until 
convergence.  As a numerical check, we obtain the same results using 
Monte Carlo integration to compute the angular integral in Eq.~(\ref{P})
by summing over randomly sampled directions.

\begin{figure}
\centering\includegraphics[width=0.5\textwidth]{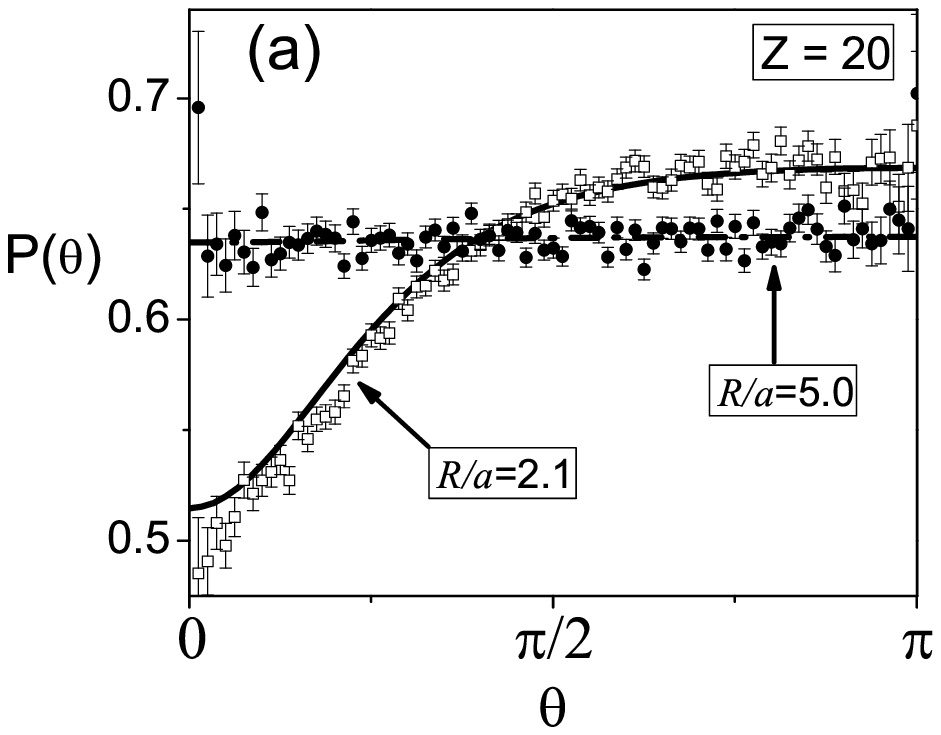}
\centering\includegraphics[width=0.5\textwidth]{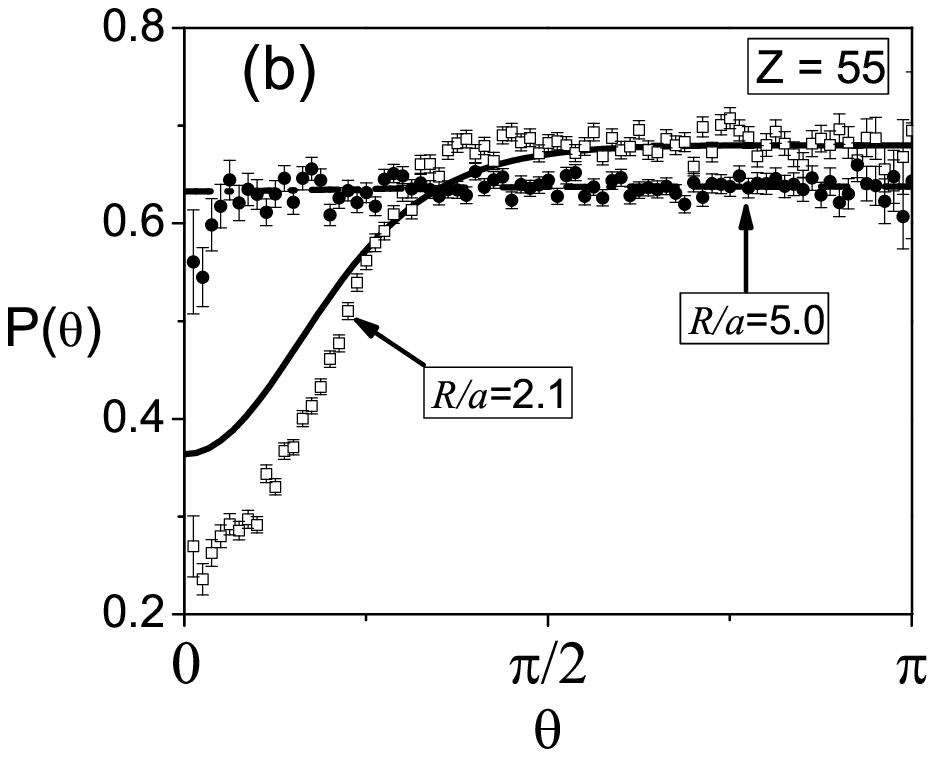}
\centering\includegraphics[width=0.5\textwidth]{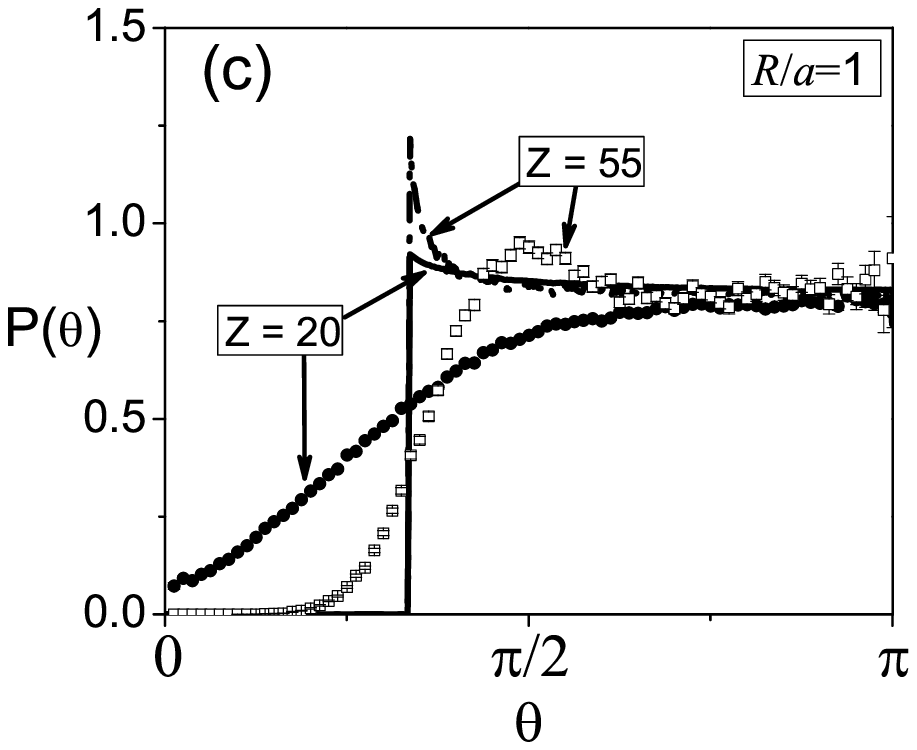}
\caption{\label{odf}{Arm orientational distribution function $P(\theta)$ 
vs. polar angle $\theta$ for two 8-arm stars of radius (arm length) $a=10$ nm, 
with $10$ beads per arm, Debye screening constant $\kappa a=1$, and 
various labeled star valences $Z$ and center-center separations $R$.
Results from Monte Carlo simulation (symbols) and density-functional 
theory (curves) are compared.  Error bars represent statistical uncertainty 
(one standard deviation) in simulation data.}}
\end{figure}

Figure~\ref{odf} compares our MC simulation results and DFT predictions 
for $8$-arm stars, each of radius $a=10$ nm and with $10$ beads per arm,
for a Debye screening constant $\kappa a=1$ and various star valences
and separations.  The error bars represent root-mean-squared deviations 
from the average of intermediate averages over $100$-cycle intervals.  
Evidently, the higher the valence and the closer the separation of the 
two stars, the more pronounced the anisotropy of the ODF, consistent with 
intuition.  In the case of nonoverlapping stars, the DFT predictions agree 
well with simulation for a relatively low valence ($Z=20$), as seen in 
Fig.~\ref{odf}(a).  For a higher valence ($Z=55$), the mean-field theory's 
neglect of arm-arm correlations leads to an underestimation of the ODF 
structure [Fig.~\ref{odf}(b)].  

\begin{figure}
\centering\includegraphics[width=0.8\textwidth]{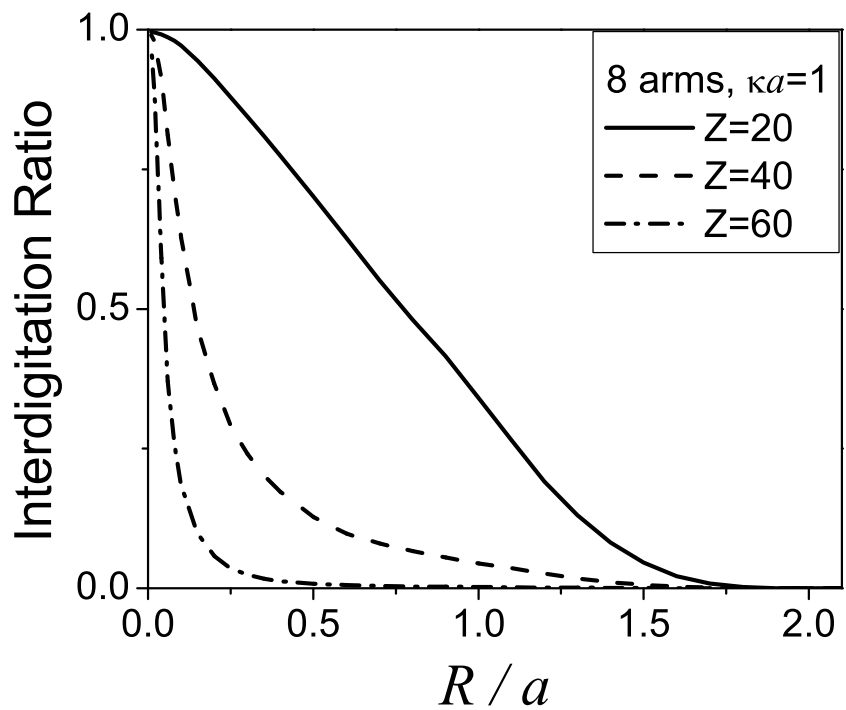}
\caption{\label{interdigit}{Fraction of interdigitating configurations 
(see text) for two overlapping 8-arm stars of various valences at 
fixed screening constant $\kappa a=1$.}}
\end{figure}

In the case of overlapping stars, the DFT's simplifying assumption of 
no interdigitation qualitatively models suppression of the ODF in the 
forward (small-$\theta$) direction, but the sharp cut-off in $P(\theta)$
is clearly too severe.  As Fig.~\ref{odf}(c) shows, Brownian motion 
at nonzero temperature can cause the forward arms of two overlapping stars 
to partially interdigitate.  
To quantify this behavior, we define an interdigitating configuration 
as one in which an arm on one star pierces a triangle whose vertices are
the tips of two arms on the second star and the center of that star.  
Based on this criterion, we define the ``interdigitation ratio" as the 
number of accepted configurations with interdigitating arms divided by
the total number of accepted configurations.  As Fig.~\ref{interdigit} 
shows, the interdigitation ratio increases with decreasing star separation 
and with decreasing star valence.  
Nevertheless, the extent of interdigitation is sufficiently weak that 
the DFT's neglect of such configurations still yields an accurate 
effective pair potential (see below).

\begin{figure}
\centering\includegraphics[width=0.5\textwidth]{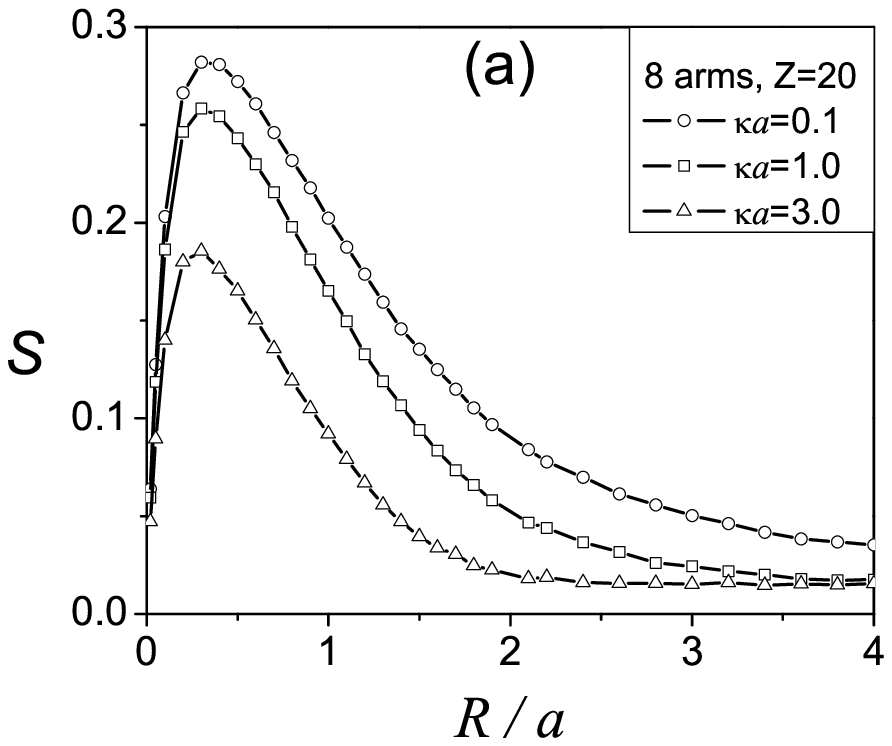}
\centering\includegraphics[width=0.5\textwidth]{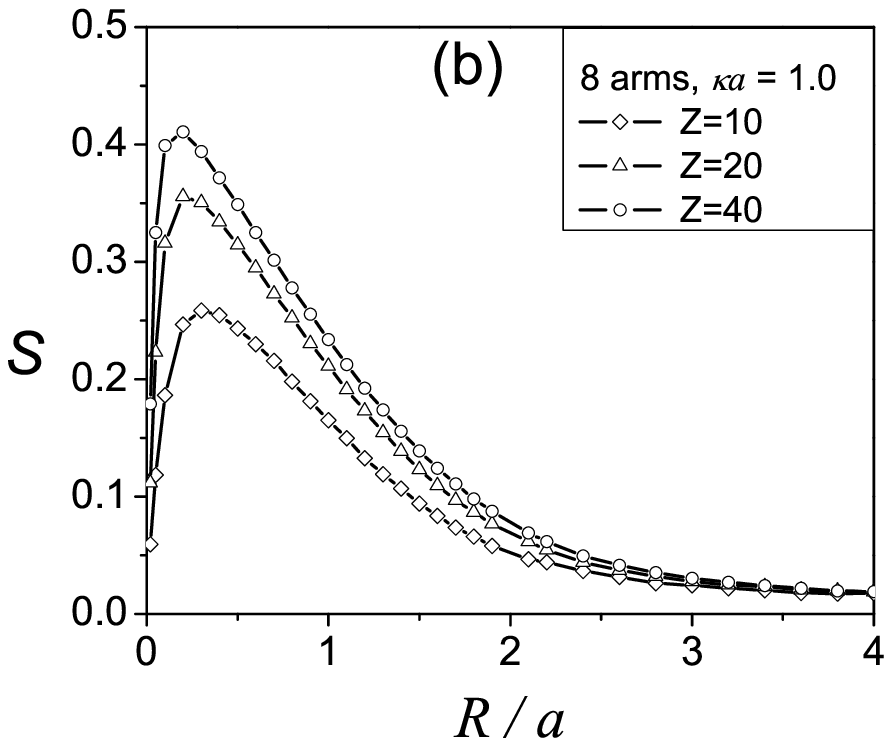}
\centering\includegraphics[width=0.5\textwidth]{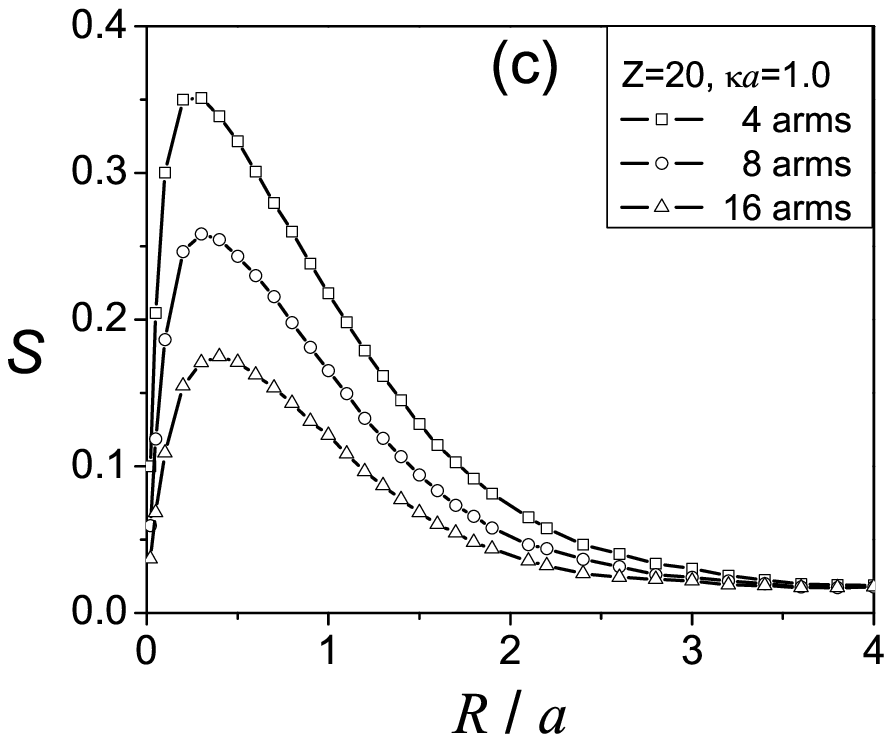}
\caption{\label{oop}{Orientational order parameter $S$ vs.
center-center separation $R$ between two stars of radius $a=10$ nm,
with 10 beads per arm, for various parameter combinations. 
(a) Fixed arm number and star valence, varying screening length. 
(b) Fixed arm number and screening length, varying star valence. 
(c) Fixed screening length and valence, varying arm number.
Symbols are simulation data and curves are guides to the eye.}}
\end{figure}

To further quantify the extent of anisotropy in the arm orientational 
distribution, we define an orientational order parameter 
\begin{equation}
\label{orderP} S=-\langle\cos\theta\rangle
=-\frac{2\pi}{f}\int^{\pi}_{0}{\rm d}\theta\sin\theta
\cos\theta P(\theta).
\end{equation}
For a perfectly isotropic distribution, $S=0$, whereas for a highly 
anisotropic distribution for which the ODF peaks at $\theta=\pi$, $S \cong 1$.  
Extracting the order parameter from a simulation simply amounts to 
computing the sum
\begin{equation}
\label{orderPMC} S=-\frac{1}{\sum_{k=1}^{M}N_{k}}\sum_{k=1}^{M} 
N_{k}\cos\theta_{k}.
\end{equation}
Figure~\ref{oop} displays the orientational order parameter as calculated 
from simulation for various choices of parameters, namely star valence, 
arm number, and screening length.  As seen in Figs.~\ref{oop}(a) and (b), 
for fixed arm number the order parameter increases with increasing star 
valence and increasing screening length (decreasing screening constant).  
Figure~\ref{oop}(c) demonstrates that for fixed valence and screening length
the order parameter increases upon reducing the number of arms (increasing
charge per arm).  Thus, orientational ordering (anisotropy) is amplified 
by strengthening or increasing the range of electrostatic interactions,
which is consistent with the trends observed in the ODF (Fig.~\ref{odf}).  

\begin{figure}
\centering\includegraphics[width=0.5\textwidth]{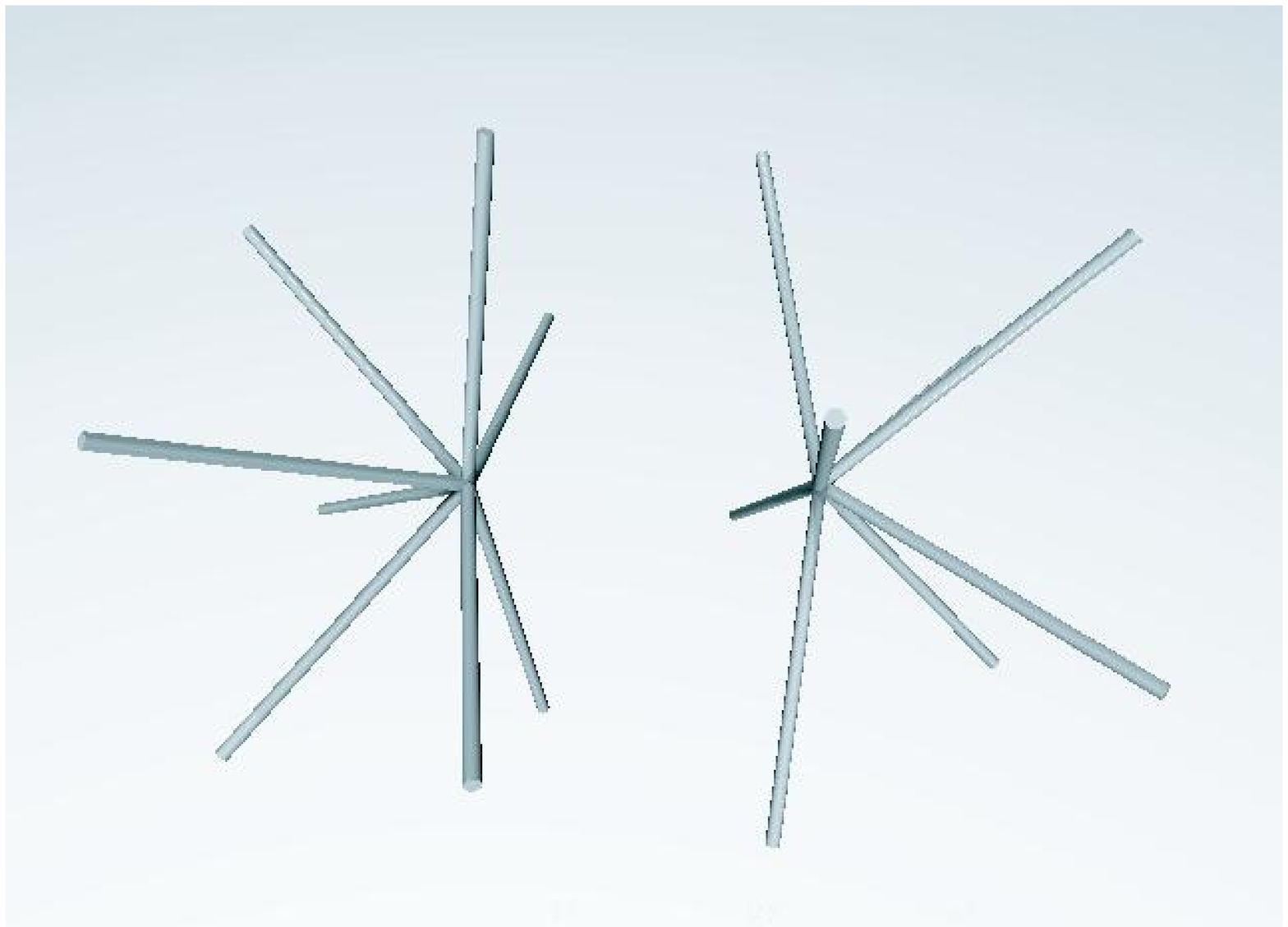}
\centering\includegraphics[width=0.5\textwidth]{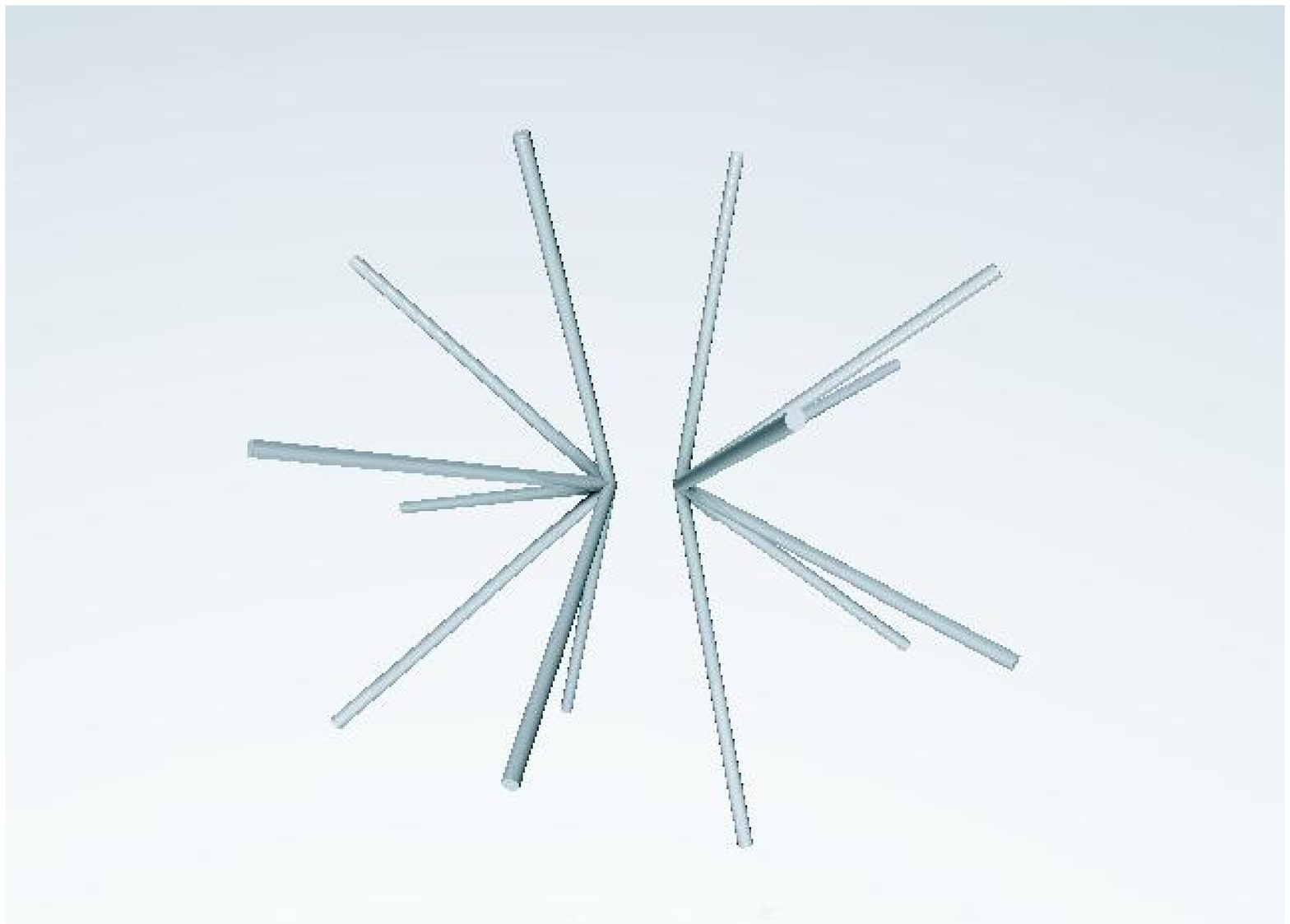}
\caption{\label{snapshot}{Configurations of two approaching
stars at center-center distance $R/a=1$ (upper panel) and
$R/a=0.2$ (lower panel) from a torque balance analysis.  
Each star has $8$ arms of length $10$ nm, with 10 beads per arm,
valence $Z=20$, and screening constant $\kappa a=1$.}}
\end{figure}

Interestingly, the degree of anisotropy does not vary monotonically with 
center-center distance between two stars.  On the contrary, it attains 
a maximum at a distance where two stars strongly overlap, but short of 
complete overlap.  The abrupt decrease in $S$ at very short separations 
simply reflects the loss of identity of completely overlapping stars that 
merge into one isotropic ``super-star."  As an aid to visualizing the 
anisotropy, Fig.~\ref{snapshot} provides snapshots of the arm orientations 
of two 8-arm stars, as determined from the torque balance analysis and 
MC simulation.

\begin{figure}
\centering\includegraphics[width=0.6\textwidth]{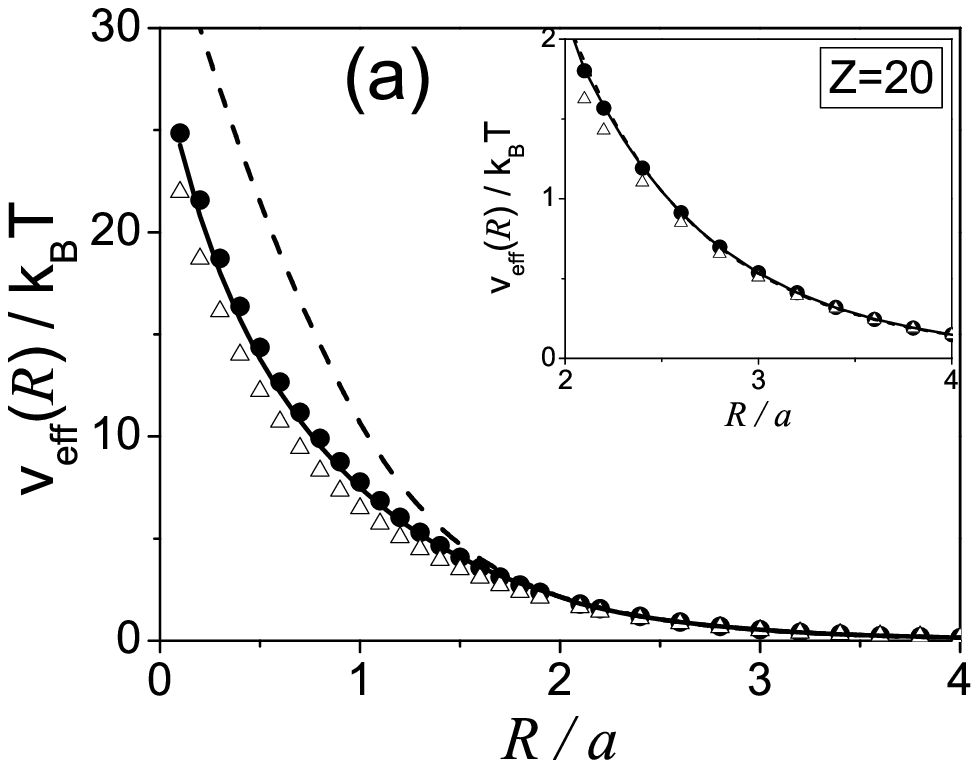}
\centering\includegraphics[width=0.6\textwidth]{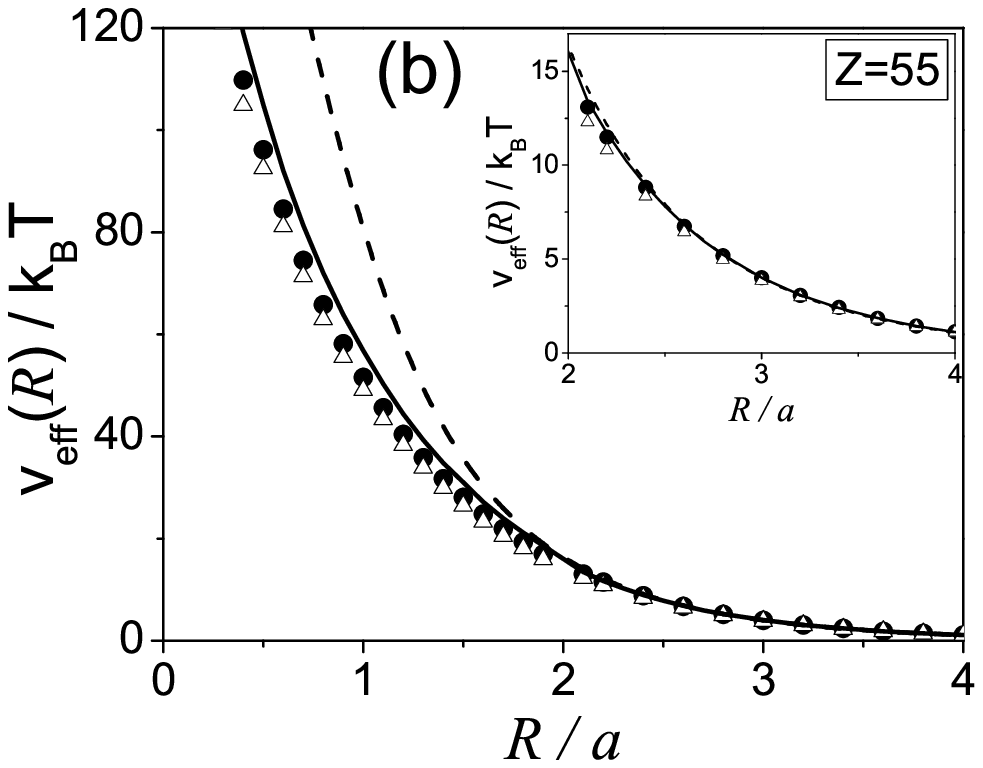}
\caption{\label{veffr}{Effective pair potential vs.
center-center separation between an isolated pair of 8-arm stars of
valence (a) $Z=20$ and (b) $Z=55$, calculated from Monte Carlo
simulation ($\bullet$), density-functional theory (solid line),
torque balance analysis ($\bigtriangleup$), and linear response
theory+continuum model~\cite{Denton03} (dashed line).
Insets magnify nonoverlapping region.}}
\end{figure}

Next, we compare results for the effective star-star pair potential, 
as computed from MC simulation, torque balance analysis, and DFT 
[Eqs.~(\ref{energyMC})-(\ref{eqnewtonian}), (\ref{P}), and (\ref{EnergyDFT})].
As seen in Fig.~\ref{veffr}, the torque balance analysis predicts the 
lowest pair energy of the three methods, as expected of a method designed
to find the ground state configuration.  The actual energy is higher
because of thermal rotational motion of the arms at nonzero temperature.  
The DFT predictions closely track the simulation data, aside from 
relatively small deviations at overlapping separations.  

The consistent agreement between theory and simulation at all separations 
is somewhat unexpected, considering the significant discrepancies in arm 
orientational distributions for high valences and for overlapping stars 
[Fig.~\ref{odf}(c)].  The theory's complete neglect of arm correlations 
and interdigitation is obviously unphysical for significantly overlapping 
stars, especially in the forward direction, {\it i.e.}, for small $\theta$.  
As a consequence, the theory underestimates the density of (interdigitating) 
arms within the forward cone [Fig.~\ref{overlap}] and overestimates the 
arm density just outside of the forward cone.  For stars that are only 
weakly overlapping, the ODF discrepancies are limited to arm configurations 
with small polar angles, which account for a relatively low fraction of 
the phase space in the angular integral for the effective pair potential 
[Eq.~(\ref{EnergyDFT})].  For more strongly overlapping stars, the close 
agreement between theory and simulation must be attributed to a fortuitous 
cancellation of errors -- namely an overestimate of the contribution to
$v_{\rm eff}(R)$ from arms outside of the interdigitation cone compensated 
by an underestimate of the contribution from interdigitating arms.
Although it should be feasible to devise an analytical fit to our results 
for the effective pair potential of two overlapping stars for any set of 
system parameters, namely star radius, arm number, linear charge density,
and screening length, we leave this fitting analysis to a future study.

\begin{figure}
\centering\includegraphics[width=0.8\textwidth]{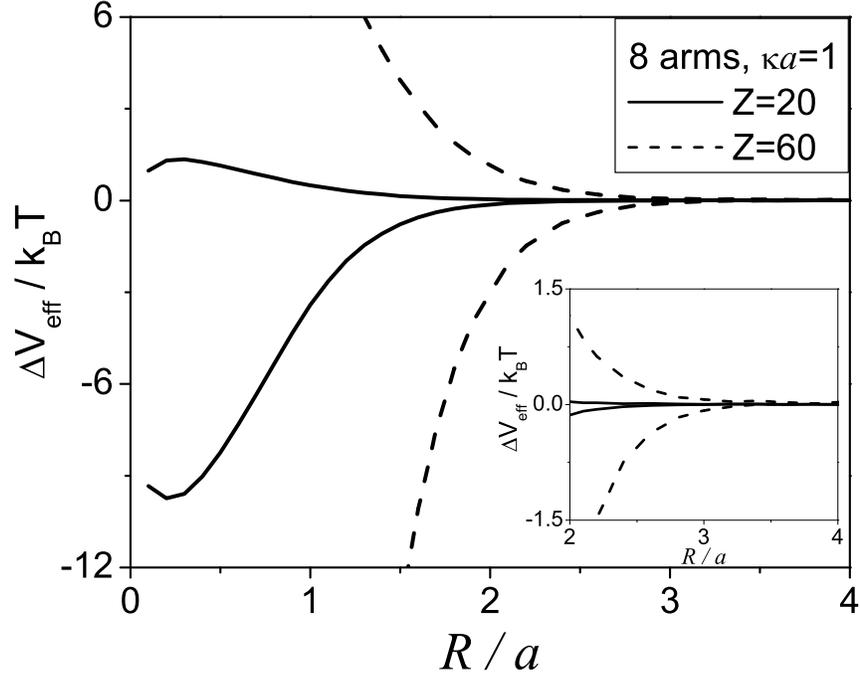}
\caption{\label{delta-veff}{Monte Carlo simulation results for 
anisotropy-induced changes in inter- and intra-star pair interaction 
energies for $8$-arm stars with arm length $a=10$ nm, 10 beads per arm,
and screening constant $\kappa a=1$.
Solid and dashed curves are intra-star (upper) and inter-star (lower) 
energies for star valences $Z=20$ and $Z=60$, respectively.  
Inset magnifies nonoverlapping region.}}
\end{figure}

For nonoverlapping stars, the simulation and DFT results hardly deviate from 
the predictions of linear response theory applied to an isotropic continuum 
model for the $1/r^2$ charge distribution of PE stars~\cite{Denton03}.  
The latter theory predicts a screened-Coulomb (Yukawa) potential between 
pairs of stars of the form
\begin{equation}
v_{\rm eff}(R)~=~\frac{Z^2e^2}{\epsilon}~\left[\frac{{\rm shi}
(\kappa a)}{\kappa a}\right]^2~\frac{e^{-\kappa R}}{R}, \qquad R>2a,
\label{veff-lrt}
\end{equation}
where ${\rm shi}(x)\equiv\int_0^x{\rm d}u\,\sinh(u)/u$ denotes the 
hyperbolic sine integral function.
The accuracy of Eq.~(\ref{veff-lrt}) may seem surprising, given the 
anisotropic arm distributions (Fig.~\ref{odf}).  Closer examination reveals, 
however, that the anisotropy-induced reduction of inter-star interaction 
energy is largely balanced by an increase of the intra-star interaction 
energy, as illustrated in Fig.~(\ref{delta-veff}). 

\begin{figure}
\centering\includegraphics[width=0.8\textwidth]{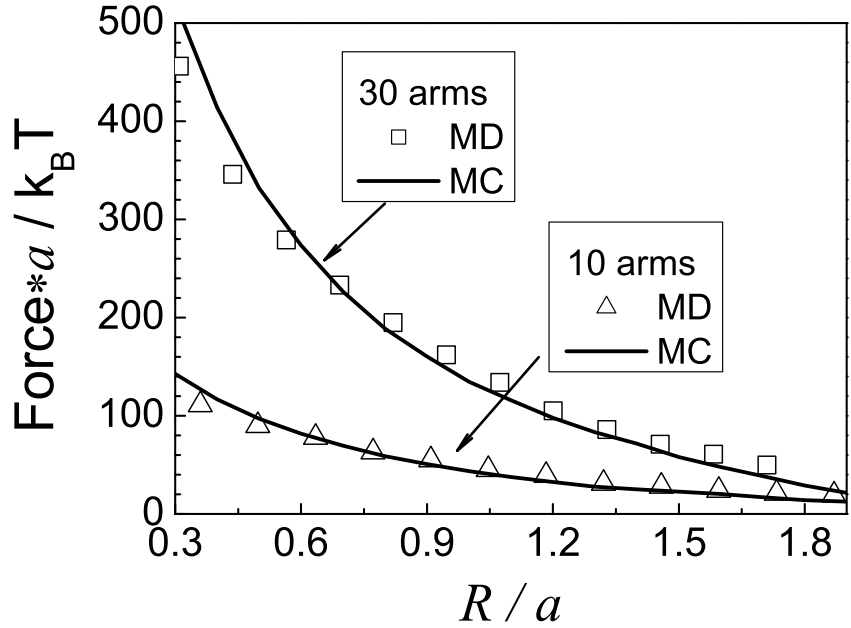}
\caption{\label{feff}{Comparison of forces between two stars 
of radius $a$ vs. center-center separation $R$ from our Monte Carlo (MC)
simulations of the rigid-arm-Yukawa model (curves) and molecular dynamics (MD)
simulations of a molecular model (symbols)~\cite{Jusufi-jcp02}.  
Choosing the number of condensed counterions as $N_1=137$ for $10$-arm 
stars and $N_1=465$ for $30$-arm stars yields good agreement 
between MC and MD data (see text).}}
\end{figure}

Finally, as a test of the validity of the rigid-arm-Yukawa model and the 
accuracy of our methods, we compare with results from molecular dynamics 
simulations of a 
molecular model of PE stars~\cite{Jusufi-prl02,Jusufi-jcp02} in which the 
arms are represented as bead-spring chains of Lennard-Jones particles 
connected by a finite-extension-nonlinear-elastic (FENE) potential and 
charged monomers and counterions interact via bare Coulomb potentials.
The strong stretching of chains observed in the MD 
simulations~\cite{Jusufi-prl02,Jusufi-jcp02} allows direct comparison with 
our simpler rigid-arm model.  To facilitate comparison, we choose parameters 
to match those given in Table IV of Jusufi {\it et al.}~\cite{Jusufi-jcp02}, 
including the Debye screening length, Bjerrum length $e^2/\epsilon 
k_{\rm B}T=0.75$ nm, and fraction of charged beads per arm $\alpha$.  
Following refs.~\cite{Jusufi-prl02,Jusufi-jcp02}, we treat the effective 
star valence $Z$ as a tunable parameter to fit the effective pair force 
$F=-\partial v_{\rm eff}(R)/\partial R$.  (Note that tuning $Z$ also 
changes the screening length.)  From the effective valence, we then obtain
the effective number of counterions condensed inside the two stars,
$N_1=2(fN_b\alpha-Z)$.  As shown in Fig.~\ref{feff}, our best fits to the 
MD forces between overlapping pairs of stars with charging fraction 
$\alpha=1/4$ are obtained by choosing $N_1=137$ for 10-arm stars and 
$N_1=465$ for 30-arm stars.  For comparison, the corresponding MD 
values~\cite{Jusufi-note} are, respectively, $N_1=147$ and $N_1=450$.
The excellent fit of the MC and MD force data, in particular the close
agreement of the effective number of condensed counterions, provides a
consistency check and supports the simple rigid-arm-Yukawa model for the
weakly charged stars considered here.  Further such comparisons would be
useful to establish the limits of validity of the model.

Three issues merit further discussion.  First, while molecular models 
that include chain flexibility and explicit microions naturally provide 
the most realistic description of PE stars, the many degrees of freedom 
present major computational challenges.  Simulations of such models 
have thus far been limited to only one or two stars~\cite{RogerDelsanti04,
Jusufi-prl02,Jusufi-jcp02}.  By comparison, the relatively coarse-grained 
model explored here seems to accurately capture effective interactions 
between PE stars, yet is simple enough to allow studies of phase behavior 
in simulations of many interacting stars.  

Second, although simulations of the coarse-grained and molecular models 
yield very similar pair forces, the rather different treatments of the 
microions in the two models suggest complementary physical interpretations.  
In the molecular model, the interactions between overlapping stars 
are attributed largely to entropy of trapped counterions, electrostatics 
playing only a relatively minor role~\cite{Jusufi-prl02,Jusufi-jcp02}.
In contrast, in the coarse-grained one-component model, where the microions 
are integrated out in a first step to obtain effective (screened-Coulomb)
arm-arm interactions, electrostatics is the sole determiner of effective 
star-star interactions.  Thus, the entropy of trapped counterions, which
in the molecular model is explicitly associated with the distribution of
counterions around the arms of the stars, has its coarse-grained counterpart 
in the implicit contribution of counterions to the effective star valence 
and to the effective interactions via the Debye screening constant.  
Although the correspondence between the two models has been tested here 
using the available MD data for 10-arm and 30-arm stars [Fig.~\ref{feff}], 
it should be tested more extensively in future, especially over a broader
range of star functionality.

Third, the PE stars considered here belong to a class of macromolecules 
for which chain entropy is negligible.  Since we model the arms as charged 
rigid rods, and ignore excluded-volume interactions between arms, the 
effective pair interactions vanish in the limit of charge-neutral stars.  
For stars with flexible arms, Witten and Pincus~\cite{WittenPincus86} and 
Likos {\it et al.}~\cite{LikosRichter98} have shown that conformational 
entropy of the arms generates an effective ``ultrasoft" repulsion between 
neutral stars, which varies logarithmically with distance at small separations 
and follows a Yukawa form at longer separations.  
Such entropic interactions are here neglected.

\section{Summary and Conclusions}\label{Conclusions}

In summary, within a coarse-grained model of polyelectrolyte stars, we have 
analyzed arm orientational anisotropy and effective interactions between
an isolated pair of weakly charged (orientationally molten) stars.  
The arms of the 
stars were modeled as rigid rods, freely rotating about the star centers, 
and carrying evenly spaced charged beads.  Microions were included implicitly 
by modeling electrostatic interactions between pairs of charged beads 
via an effective screened-Coulomb (Yukawa) potential.  Arm orientational 
distribution functions and effective star-star electrostatic interactions 
were calculated by three independent methods: Monte Carlo simulation, 
torque balance analysis, and density-functional theory.

The arm orientational distributions predicted by DFT are in good agreement 
with simulation for nonoverlapping stars of relatively low effective valence.
With increasing overlap and increasing valence, the agreement worsens 
due to the theory's neglect of arm-arm correlations and interdigitation.
An orientational order parameter, quantifying the extent of anisotropy, 
increases as two stars approach each other, peaks at strong overlap, and 
then decreases as the stars near complete overlap.

The simulation and DFT results for the effective pair potential between 
nonoverlapping stars are in excellent agreement with previous predictions 
of linear response theory applied to a continuum model of PE stars with 
isotropic $1/r^2$ charge distribution~\cite{Denton03}.  
This close agreement results not only from the relatively weak anisotropy 
of separated stars, but also from a fortuitous cancellation between reduced
inter-star interactions and enhanced intra-star interactions induced by 
arm anisotropy.  We conclude that electrostatic interactions between 
nonoverlapping weakly charged PE stars can be accurately modeled by a 
simple Yukawa effective pair potential.

The effective pair potentials from our Monte Carlo simulations and DFT 
calculations are in good agreement over the whole range of star separations, 
including strongly overlapping stars, for the parameters here investigated.  
Furthermore, treating the effective valence of the stars as a 
fitting parameter, we achieve a close fit to available data for the 
effective pair forces and numbers of condensed counterions from 
MD simulations~\cite{Jusufi-prl02,Jusufi-jcp02} of a molecular model 
with flexible bead-spring chains and explicit counterions.

Simulations of many-star systems are a natural next step.  For dilute
solutions, interactions between isolated pairs of nonoverlapping stars 
can be reliably approximated, as confirmed here, by an isotropic 
screened-Coulomb pair potential~\cite{Denton03}.  Only interactions 
between overlapping stars must then be computed by explicit summation 
over pairs of charged beads, pending an analytical fit to our MC data.  
For concentrated solutions of frequently overlapping stars, arm anisotropy 
and many-body interactions among three or more stars are likely to be 
important.  In this regime, explicit pairwise summation over beads is 
essential.  Such an approach could independently test the phase behavior 
predicted by recent simulations of dense solutions of stars interacting 
via isotropic effective pair potentials~\cite{Hoffmann04}.

\acknowledgements
We thank Ben Lu, Shrikant Shenoy, and Christos N.~Likos for helpful 
discussions.  This work was supported by the National Science Foundation 
under Grant No. DMR-0204020.


\newpage




\end{document}